%% file: main.tex
\documentclass[conference]{IEEEtran}
\IEEEoverridecommandlockouts

\usepackage{cite}
\usepackage{amsmath,amssymb,amsfonts}
\usepackage{graphicx}
\usepackage{textcomp}
\usepackage{xcolor}
\usepackage{algorithm}
\usepackage{algpseudocode}
\usepackage{booktabs}
\usepackage{array}
\usepackage{multirow}
\usepackage[hidelinks]{hyperref}
\usepackage{orcidlink}
\usepackage{pbalance}
\usepackage{soul}

\newcommand{\E}{\mathbb{E}}
\newcommand{\cvar}{\mathrm{CVaR}}
\newcommand{\eens}{\mathrm{EENS}}
\newcommand{\lolh}{\mathrm{LOLH}}
\newcommand{\Pmax}{P^{\max}}
\newcommand{\Emax}{E^{\max}}

\begin{document}
\title{Storage as a Transmission Asset (SATA) for Large-Load Congestion Relief \vspace{-0.25em}}
\bstctlcite{IEEEexample:BSTcontrol}

\author{%
\IEEEauthorblockN{%
Abanish~Tiwari\orcidlink{0009-0003-0609-8571},
Chandan~Chaudhary\orcidlink{0009-0002-2389-9568},~\emph{Student Member, IEEE},
Yansong~Pei\orcidlink{0000-0002-4647-7491},~\emph{Member, IEEE},\\
Mohammed~Ben-Idris\orcidlink{0000-0002-8731-8913},~\emph{Senior Member, IEEE},
and Joydeep~Mitra\orcidlink{0000-0001-9287-0983},~\emph{Fellow, IEEE}%
}
\IEEEauthorblockA{%
Electrical and Computer Engineering, Michigan State University, East Lansing, MI 48824, USA\\
E-mails: tiwariab@msu.edu;\; chaud152@msu.edu;\; peiyanso@msu.edu;\; benidris@msu.edu;\; mitraj@msu.edu%
}
\vspace{-2.4em}
}

\maketitle

\begin{abstract}
%Growth in hyperscale data centers and other large concentrated loads can create a transmission-capacity shortfall when new demand exceeds the transfer capability of existing import corridors. In systems with thermally constrained corridors, this shortfall can lead to load curtailment, delayed interconnection, or the need for network reinforcement. 
Hyperscale data centers and other large concentrated loads can impose substantial new demand on existing transmission networks. If import corridors lack sufficient transfer capability, operators may need to curtail load, delay interconnection, or reinforce the network to maintain reliable service.
An energy storage system (ESS) deployed as a storage-as-transmission asset (SATA) offers a non-wires alternative by providing operator-directed support to constrained import corridors. However, the operating-level reliability value of SATA dispatch remains insufficiently quantified. This paper evaluates operator-directed SATA using a day-ahead DC optimal power flow that co-optimizes generation, ESS dispatch, and load curtailment across Monte Carlo scenarios of demand and generator availability. Operating reliability is assessed using expected energy not served (EENS), loss-of-load hours (LOLH), and the conditional value at risk (CVaR) of daily unserved energy. Congestion-price and flow-sensitivity metrics are used to identify the limiting corridor and storage location. The interconnection is then screened to determine whether SATA is suitable, reinforcement is required, or storage would provide little transmission value. Results show that operator-directed SATA reduces average unserved energy, loss-of-load exposure, and tail risk compared with deploying the same ESS for pure arbitrage. These results demonstrate that the operating designation of storage is a primary driver of its transmission value.

\end{abstract}

\begin{IEEEkeywords}
conditional value-at-risk (CVaR), congestion relief, data centers, energy storage system (ESS), large-load, reliability, storage as transmission asset (SATA).
\end{IEEEkeywords}

% =====================================================================
\section{Introduction}
% =====================================================================
\input{Contents/introduction}

% =====================================================================
\section{System Model and Problem Formulation}
\enlargethispage{2\baselineskip}
\label{sec:model}
% =====================================================================
\input{Contents/methodology}

% =====================================================================
\vspace{-1em}
\section{Test System and Results}
\label{sec:results}
% =====================================================================
\input{Contents/results}

% =====================================================================
\section{Conclusion}
\label{sec:conclusion}
% =====================================================================
\input{Contents/conclusion}

\balance
% \enlargethispage{1\baselineskip}
\bibliographystyle{IEEEtran}
\bibliography{Contents/References}
\end{document}

%% file: Contents/introduction.tex
% Introduction
Rapid growth in large, concentrated loads, especially AI data centers, is creating new challenges for transmission systems. Data center electricity consumption is projected to represent 6.7--12\% of national electricity demand by 2028 \cite{Shehabi2024}. In the United States, transmission infrastructure often requires several years, and in some cases nearly a decade, to plan, permit, and construct, whereas large loads are typically developed and interconnected within two to three years. Thermal limits on existing corridors therefore delay load interconnection and constrain system adequacy.
\enlargethispage{2\baselineskip}

Energy storage systems (ESS) deployed at a thermally constrained corridor can provide temporary congestion relief. The ESS charges during off-peak hours and discharges during peak demand, which reduces corridor loading and raises the transferable power capacity of the existing infrastructure. This capability allows a large load to be served on the existing corridor while long-term transmission upgrades are planned or constructed.

The regulatory framework for this application has been established in recent years \cite{Twitchell2024}. Storage operated as a transmission asset (SATA) is dispatched under the direction of the transmission operator, maintains a sufficient state of charge to perform its transmission function, and recovers costs through regulated transmission rates~\cite{Twitchell2024}. By contrast, storage operated in place of a transmission asset (SIPTA) participates in electricity markets and recovers costs through market revenues, while dual-use storage combines both functions \cite{Twitchell2024, osti_3013844}. Transmission-only storage tariffs filed by MISO, ISO-NE, and SPP between 2020 and 2023 restrict these designated assets to transmission service and reinforce the regulatory separation between SATA and market resources \cite{FERC841}.

Storage for congestion relief and transmission support has been studied widely. Early work addressed congestion relief and storage siting on thermally limited paths \cite{DelRosso2014,Pandzic2015}. Subsequent studies formulated storage as a transmission alternative in planning \cite{Arteaga2021} and assessed its reliability contribution under generation variability \cite{Sulaeman2017}. Although prior studies demonstrate the value of storage on constrained corridors, their day-ahead reliability implications for large loads served through those corridors remain insufficiently quantified.

Energy storage has also been examined as a reliability and resilience resource in renewable-rich and weather-exposed systems. This body of work covers renewable variability mitigation \cite{sanchez2024EnergyStorage,Chaudhary2025Quantitative}, distributed storage coordination during extreme-weather events \cite{chaudhary2025resilience,chaudhary2026adaptive}, and peak-demand control with machine learning \cite{11562943}. Such studies also establish storage as a flexible reliability resource. However, limited attention has been given to the operating strategy that governs whether an ESS serves as a regulated transmission asset or participates as a market resource.
\enlargethispage{2\baselineskip}

For transmission-service applications, energy storage has been dispatched as virtual transmission in deterministic day-ahead operation \cite{7540964, WangLi2023}. At the planning level, previous studies have evaluated the reliability contribution of storage using forced-outage simulations \cite{Sehloff2025} and demonstrated that its effectiveness as a transmission substitute is fundamentally constrained by its finite energy capacity \cite{Zhang2024TVES}. Dynamic line rating provides a complementary non-wires option, and the joint deployment of these two resources is examined in \cite{Tiwari2026DLR}.

Three gaps emerge from the existing literature. First, the SATA literature focuses primarily on planning decisions for siting, sizing, and operating-condition analysis \cite{Pandzic2015,Arteaga2021,Sehloff2025}, or on deterministic day-ahead dispatch without stochastic uncertainty \cite{WangLi2023}. An operating-level reliability assessment against the adequacy indices that transmission operators use has not been reported. Second, the same ESS can be operated either as a SATA or as a pure arbitrage resource, yet the reliability of each operating strategy has not been evaluated separately. Third, the ESS bus and rating are typically assumed in the SATA literature rather than derived from system parameters. The existing literature gives limited attention to determining whether a given interconnection is suitable for storage-based congestion relief or instead requires conventional network reinforcement. The slow uptake of SATA has been attributed in \cite{Twitchell2024} to exactly this absence of concrete, operating-level evidence.

This paper addresses these gaps with three contributions. \emph{First}, a day-ahead reliability assessment quantifies the adequacy impact of operator-dispatched SATA for a large load and reports performance with expected energy not served (EENS), loss-of-load hours (LOLH), and the conditional value at risk (CVaR) tail of daily unserved energy. \emph{Second}, the ESS bus and power rating are derived from congestion-price signals, and a companion criterion screens whether storage can relieve a given interconnection or only reinforcement will suffice. \emph{Third}, the same hardware is dispatched as a SATA and as a pure-arbitrage resource on identical scenarios, which isolates the reliability contribution of the operating designation from the physical capability of the ESS.

The rest of the paper is organized as follows. Section~\ref{sec:model} presents the network model, day-ahead dispatch formulation, ESS model, siting and sizing rule, scenario model, and reliability metrics. Section~\ref{sec:results} reports the case study and results. Section~\ref{sec:conclusion} concludes.

%% file: Contents/methodology.tex
% System Model and Problem Formulation
This section develops the network model, day-ahead optimal power flow, ESS model, siting and sizing rule, scenario model, and reliability metrics used to evaluate three operating cases: no storage, storage dispatched for pure energy arbitrage, and storage operated as a SATA.

\subsection{Network and shift factors}
The network uses the standard linear (DC) model \cite{wood2013power}. The power transfer distribution factor (PTDF) matrix $H$ maps bus injections to line flows,
\begin{equation}
f_\ell=\sum_{i\in\mathcal B}H_{\ell,i}\,p_i ,
\label{eq:ptdf}
\end{equation}
where $\mathcal B$ is the set of buses, $p_i$ is the net injection at bus $i$, and the reference bus absorbs the balance. Through~\eqref{eq:ptdf}, an injection at the ESS bus reduces flow on the constrained corridor, which constitutes the primary congestion-relief mechanism utilized by the dispatch.

\subsection{Day-ahead DC optimal power flow}
For each day the operator solves a cost-minimizing DC optimal power flow over the hours $t\in\mathcal T$, with generator outputs $P_{g,t}$ at marginal cost $C_g$, demand $D_{i,t}$, load curtailment $s_{i,t}$ priced at the value of lost load $V$, and the ESS at bus $k^\star$:
\begin{subequations}\label{eq:opf}
\begin{equation}
\min_{P,\,s,\,c,\,d}\ \sum_{t\in\mathcal T}\!\Big[\sum_{g}C_g P_{g,t}+V\sum_{i}s_{i,t}\Big]
\end{equation}
\begin{equation}
\text{s.t.}\quad\sum_{g}P_{g,t}+d_t+\sum_{i}s_{i,t}=\sum_{i}D_{i,t}+c_t\;\;[\lambda_{t}] \label{eq:bal}
\end{equation}
\begin{equation}
p_{i,t}=\!\sum_{g\in i}P_{g,t}-D_{i,t}+s_{i,t}+(d_t-c_t)\mathbb{1}[i{=}k^\star] \label{eq:inj}
\end{equation}
\begin{equation}
\Big|\sum_{i}H_{\ell,i}\,p_{i,t}\Big|\le a_\ell\,f_\ell^{\max}\;\;[\mu_{\ell,t}] \label{eq:line}
\end{equation}
\begin{equation}
0\le P_{g,t}\le a_g\,P_g^{\max},\quad 0\le s_{i,t}\le D_{i,t}. \label{eq:bnd}
\end{equation}
\end{subequations}
Constraint~\eqref{eq:bal} is the system power balance, where $\lambda_t$ is the dual variable associated with the energy balance constraint and represents the marginal energy price. Constraint~\eqref{eq:inj} defines the nodal net injection $p_{i,t}$ that enters the shift-factor product in~\eqref{eq:ptdf}. Constraint~\eqref{eq:line} enforces the thermal limit on each line, where $f_\ell^{\max}$ is the thermal capacity of line $\ell$, $a_\ell \in \{0,1\}$ is the line availability state, and $\mu_{\ell,t} \ge 0$ on a binding limit is the congestion price, the marginal value of additional capacity on that corridor \cite{Schweppe1988}. Constraint~\eqref{eq:bnd} bounds generator output, where $P_g^{\max}$ is the installed capacity of generator $g$ and $a_g \in \{0,1\}$ is its availability state, drawn from the forced-outage model in the demand and outage scenarios subsection. Because the new load is located behind the constrained corridor, curtailment is localized, and relief of the binding limit occurs only through demand reduction in that region, which defines the unserved energy of interest.

\subsection{ESS model}
\label{sub:bess}
The ESS enters the dispatch~\eqref{eq:opf} through the net injection $d_t - c_t$ at bus $k^\star$, where $c_t$ and $d_t$ are the charge and discharge power at hour $t$. The energy balance across the day follows a standard model with charge efficiency $\eta_c$ and discharge efficiency $\eta_d$ \cite{Sioshansi2009,mongird2020}:
\begin{subequations}\label{eq:soc}
\begin{align}
& E_t=E_{t-1}+\eta_c\,c_t-d_t/\eta_d,\qquad E_0=E_{\mathrm{init}}, \label{eq:socdyn}\\
& 0\le c_t\le \Pmax,\qquad 0\le d_t\le \Pmax, \label{eq:cd}\\
& E^{\mathrm{lo}}\le E_t\le E^{\mathrm{hi}},\qquad \Emax=T_d\,\Pmax, \label{eq:elim}
\end{align}
\end{subequations}
where $E_t$ is the stored energy at the end of hour $t$, $E_{\mathrm{init}}$ is the initial state of charge, $E^{\mathrm{lo}}$ and $E^{\mathrm{hi}}$ are the lower and upper bounds of the usable state-of-charge range, and $T_d$ is the rated storage duration. Constraint~\eqref{eq:cd} caps charge and discharge power at the converter rating $\Pmax$. Constraint~\eqref{eq:elim} keeps the stored energy within the usable band and ties the energy capacity $\Emax$ to the duration and power rating. The energy limit $\Emax = T_d\Pmax$ bounds how long the ESS can sustain its output, which constrains storage as a corridor substitute under prolonged stress \cite{Zhang2024TVES}. Because the round-trip efficiency is strictly below unity, simultaneous charging and discharging is not optimal, and the model therefore remains linear \cite{Sioshansi2009}.

\enlargethispage{2\baselineskip}
\subsection{SATA versus pure arbitrage operation}
\label{sub:sata}
The three cases evaluated in this paper share the same network, OPF formulation~\eqref{eq:opf}, and ESS model~\eqref{eq:soc}. The distinction across cases arises solely from the dispatch policy applied to the ESS, which constitutes the operating variable isolated in this study.

\textit{No storage:} The ESS is absent from~\eqref{eq:opf}, so $c_t = d_t = 0$ for all $t \in \mathcal{T}$. Congestion on the corridor is relieved only through generation redispatch and load curtailment.

\begin{figure}[!htbp]
\centering
\vspace{-1em}
\includegraphics[trim=0 160 400 0cm, clip, width=0.9\columnwidth]{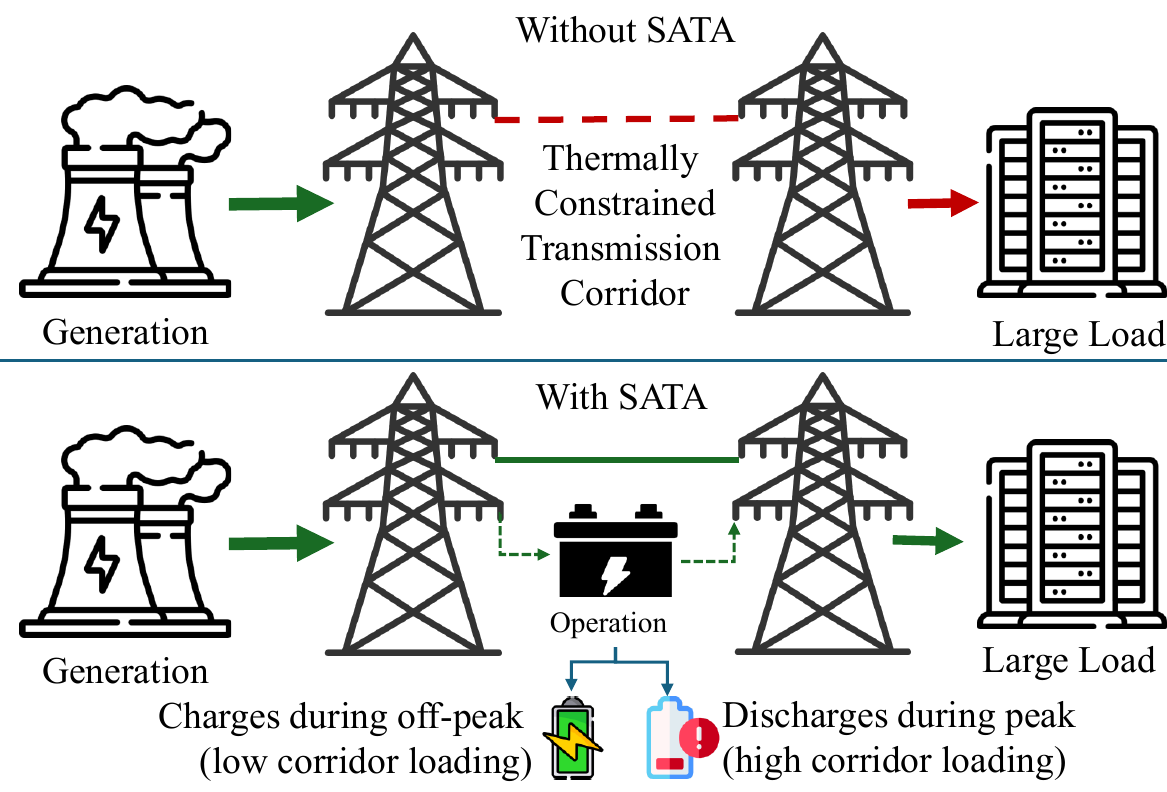}
\vspace{-1em}
\caption{Operation of ESS as SATA.}
\vspace{-1em}
\label{fig:SATA}
\end{figure}

\textit{SATA:} The charge and discharge variables $(c_t, d_t)$ are free decision variables within~\eqref{eq:opf}, co-optimized with generation and curtailment against the demand and outage scenario that is realized on that day. Because the value of lost load $V$ greatly exceeds any generation cost, the optimizer deploys the ESS first to relieve the corridor and avoid curtailment. The ESS holds energy in reserve and, when the corridor binds, discharges only the energy the transmission service requires \cite{WangLi2023,Twitchell2024,FERC841}. Under this designation, the operator is required to retain the ESS for transmission service in accordance with the regulatory framework. Fig. \ref{fig:SATA} illustrates the operating strategy of SATA. The ESS charges under light corridor loading, retains energy in reserve, and discharges during congestion to relieve thermal constraints and enhance the power transfer capability.

\textit{Pure arbitrage:} The ESS schedule $(\bar{c}_t, \bar{d}_t)$ is committed before any scenario is realized. It is computed once on the forecast day by solving
\begin{equation}
\max_{c_t,\,d_t}\ \sum_{t}\lambda_t\,(d_t-c_t)\quad\text{s.t.}\ \eqref{eq:soc},
\label{eq:arb}
\end{equation}
using forecast prices $\lambda_t$ from the day-ahead clearing \cite{Sioshansi2009}. The schedule charges during low-price hours and discharges during high-price hours of the forecast day. This committed profile is fixed in every scenario. Charging may be preempted if needed to prevent firm load curtailment, but discharge proceeds at the committed hours regardless of whether the corridor is binding at those hours. A market resource bears no obligation to hold capacity for the corridor and is not retimed to match the realized day.

The two storage cases share the same hardware, sizing, energy budget, and scenario set, and differ only in their dispatch objective and foresight. Any difference in reliability therefore reflects the operating designation, not the physical asset. This isolation is the central comparison of the paper.

\enlargethispage{2\baselineskip}
\subsection{ESS Siting, Sizing, and Applicability Screening}
\label{sub:site}

The candidate transmission corridor and ESS bus are identified from congestion-price signals derived from the day-ahead OPF \cite{DelRosso2014,Pandzic2015}. After the large load is integrated, the system is solved without the ESS over the stochastic demand scenarios, and the expected congestion price of each transmission element is computed. The target corridor $\ell^\star$ is selected as the element with the highest expected congestion price, and the ESS is located at the bus that provides the greatest sensitivity for relieving congestion on that corridor,
\begin{equation}
\ell^\star=\arg\max_{\ell}\ \E[\mu_\ell],\qquad
k^\star=\arg\max_{k}\ |H_{\ell^\star,k}|.
\label{eq:site}
\end{equation}

The expected congestion price $\E[\mu_\ell]$ identifies the corridor that most frequently limits power transfer. The PTDF coefficient $|H_{\ell^\star,k}|$ quantifies the congestion-relief effectiveness of a unit injection at bus $k$. Locating the ESS at the bus with the largest $|H_{\ell^\star,k^\star}|$ therefore maximizes corridor relief per unit of injected power.

The ESS power rating is set to offset a target fraction $\gamma$ of the limiting corridor capacity,
\begin{equation}
\Pmax=\gamma\,\frac{f^{\max}_{\ell^\star}}
{|H_{\ell^\star,k^\star}|},\qquad
\Emax=T_d\,\Pmax,
\label{eq:size}
\end{equation}
where $\gamma$ is the congestion-relief ratio and $T_d$ is the storage duration. The energy capacity follows directly from the rated duration.

The congestion analysis also screens whether storage is the appropriate tool for a given interconnection. If load curtailment occurs in the base case without the ESS, the candidate location is import-limited and requires network reinforcement rather than energy-limited storage. If no transmission constraint becomes binding, the location is uncongested, and storage provides limited congestion-relief value. The proposed SATA deployment is therefore applicable to the intermediate congested regime, where transmission constraints bind during peak loading but the system is otherwise capable of serving the demand.

\enlargethispage{\baselineskip}
% \vspace{-1em}
\subsection{Demand and outage scenarios}
\label{sub:scen}
A scenario $n$ is a day of demand together with a set of generator forced outages. Demand follows a chronological load shape scaled by bus participation factors $\phi_i$ and perturbed by a Gaussian forecast error \cite{Grigg1999}, with standard deviation
\begin{equation}
\sigma_{i,t}=\phi_i\,(L^{\max}_t-L^{\min}_t)/6 ,
\label{eq:spread}
\end{equation}
where $L^{\max}_t$ and $L^{\min}_t$ are the upper and lower bounds of the system hourly load profile at hour $t$. This six-sigma rule spans approximately $\pm 3\sigma$ of plausible demand variation \cite{Billinton1996}. The large new load is added as a near-constant block on top of this shape. Generator availability at each hour is drawn independently from the forced-outage rate,
\begin{equation}
a_g=\mathbb{1}[\xi_g>\mathrm{FOR}_g],\qquad \xi_g\sim U(0,1),
\label{eq:outage}
\end{equation}
where $\xi_g$ is a uniform random draw for generator $g$ and $\mathrm{FOR}_g$ is its forced-outage rate, taken from the test-system data \cite{Grigg1999}. The generation fleet is scaled to a representative planning reserve margin so that the binding constraint behind the corridor is the thermal import limit rather than a system-wide capacity shortage. The corridor's own forced outage is excluded because firm coverage of that contingency is the role of a parallel reinforcement rather than an energy-limited ESS. A large scenario count $N$ is required because the curtailment events that drive the adequacy indices, coincident high demand and generator unavailability behind the corridor, are infrequent.

\enlargethispage{2\baselineskip}
\subsection{Cases and metrics}
\label{sub:metric}
Each scenario is evaluated under the three cases defined in the SATA versus pure arbitrage operation subsection. The unserved energy in scenario $n$ is
\begin{equation}
U^n=\sum_{t}\sum_i s_{i,t}^n ,
\label{eq:cost}
\end{equation}
and the scenario operating cost $C^n$ is the optimal objective value of~\eqref{eq:opf}. Performance is assessed with the standard adequacy indices \cite{Billinton1996},
\begin{equation}
\eens=\frac1N\sum_n U^n,\qquad
\lolh=\frac1N\sum_n\sum_t \mathbb{1}\Big[\sum_i s_{i,t}^n>0\Big],
\label{eq:eens}
\end{equation}
where EENS is the expected energy not served per day and LOLH is the loss-of-load hours per day. Because the operator's exposure on the worst days matters as much as the average, the tail of the daily unserved energy is summarized by the conditional value at risk at level $\alpha$ \cite{RockafellarUryasev2000,Conejo2010,Poudel2019},
\begin{equation}
\cvar_\alpha(U)=\frac{1}{(1-\alpha)N}\sum_{n:\,U^n\ge q_\alpha} U^n ,
\label{eq:cvar}
\end{equation}
the mean of the worst $(1-\alpha)$ fraction of days, where $q_\alpha$ is the $\alpha$-quantile of the daily unserved energy distribution. The expected operating cost $\E[C]$ is also reported, as relieving the corridor reduces the costly redispatch the operator must otherwise pay for.

\subsection{Simulation procedure}
\label{sub:sim}
Algorithm~\ref{alg:sim} summarizes the four-step procedure.

\begin{algorithm}[!htbp]
\caption{SATA Reliability Assessment}
\label{alg:sim}
\begin{algorithmic}[1]
\Require Network topology, generator data, load profile; parameters $N$, $\gamma$, $T_d$, $\alpha$
\Ensure EENS, LOLH, $\cvar_\alpha(U)$, $\E[C]$ for no-storage, arbitrage, and SATA cases
\Statex \textit{Step A: Network setup and scenario generation}
\State Compute PTDF matrix $H$ from~\eqref{eq:ptdf}; add large load to network
\State Draw $N$ demand and outage scenarios from~\eqref{eq:spread}--\eqref{eq:outage}
\Statex \textit{Step B: Siting and sizing}
\State Solve~\eqref{eq:opf} without ESS on all $N$ scenarios; record $\E[\mu_\ell]$ for each line $\ell$
\State Determine $\ell^\star$, $k^\star$, $\Pmax$, $\Emax$ from~\eqref{eq:site}--\eqref{eq:size}
\Statex \textit{Step C: Three-case simulation}
\For{$n = 1$ \textbf{to} $N$ \textbf{(in parallel)}}
  \State No storage: solve~\eqref{eq:opf} without ESS; record $U^n_{\mathrm{NS}}$, $C^n_{\mathrm{NS}}$
  \State Arbitrage: solve~\eqref{eq:arb} on forecast; fix $(\bar{c}_t,\bar{d}_t)$; solve~\eqref{eq:opf}; record $U^n_{\mathrm{ARB}}$
  \State SATA: solve~\eqref{eq:opf} with $(c_t,d_t)$ free; record $U^n_{\mathrm{SATA}}$, $C^n_{\mathrm{SATA}}$
\EndFor
\Statex \textit{Step D: Index computation}
\State Compute EENS, LOLH from~\eqref{eq:eens} and $\cvar_\alpha(U)$ from~\eqref{eq:cvar} for each case
\end{algorithmic}
\end{algorithm}

\enlargethispage{2\baselineskip}

%% file: Contents/results.tex
% Test System and Results
This section applies the framework to the IEEE RTS-24 test system and reports reliability, risk, and cost results for the three operating cases.

\subsection{Test System and Implementation}
The framework is demonstrated on the single-area IEEE 24-bus reliability test system (RTS-24): 24 buses, 38 lines, 33 generators, a peak load of 2850~MW, and 3405~MW of installed generation capacity \cite{Grigg1999}. A 500~MW data-center load is added at Bus~3 (138~kV). Generation capacities are scaled up by 15\% so that the binding constraint behind the corridor is the thermal import limit rather than a system-wide capacity shortage. The constrained corridor, ESS bus, and ratings are then determined from~\eqref{eq:site}--\eqref{eq:size}.

The value of lost load is set to \$9000/MWh, reflecting the high interruption cost associated with data center load. The congestion-relief ratio is $\gamma = 0.15$ and the storage duration is $T_d = 4$~h \cite{mongird2020}. Charge and discharge efficiencies are both 95\%, and the state of charge is constrained to $[10\%, 90\%]$ of the energy capacity \cite{mongird2020}. The initial state of charge $E_{\mathrm{init}}$ is set to 50\%. The CVaR confidence level is $\alpha = 0.95$, corresponding to the mean of the worst 5\% of days \cite{RockafellarUryasev2000}. The scenario count is $N = 10^5$ independent daily scenarios. Because each scenario is independent, all $N$ OPF problems were solved in parallel on a desktop computer with 16~GB of RAM.

\subsection{Congestion-Price Siting and Screening}
\label{sub:endo}
The siting rule in~\eqref{eq:site}--\eqref{eq:size} is applied at three representative buses to illustrate the screening criterion. At Bus~3 (138~kV), the import transformer (3,24) carries the highest expected congestion price ($\E[\mu]\!\approx\!60$~\$/MWh) while the load is served without curtailment. Bus~3 also has the largest PTDF onto that branch ($|H|\!\approx\!0.37$). It is therefore selected as the ESS bus, and~\eqref{eq:size} gives $\Pmax\!\approx\!161$~MW and $\Emax\!\approx\!643$~MWh. At weakly connected radial buses~6 and~8, a 175~MW import line cannot serve a 500~MW load, so curtailment occurs in nearly every hour; these buses are import-limited and require network reinforcement rather than an ESS. At the strong 230~kV Bus~15, the load causes no congestion and storage provides negligible transmission benefit. The siting rule thus places the ESS at Bus~3 and correctly identifies the two regimes where storage is not the appropriate remedy: import-limited buses that need reinforcement and uncongested buses where no transmission constraint exists.

\enlargethispage{2\baselineskip}

\subsection{Results}
\label{sub:res}
Table~\ref{tab:res} compares the three cases on identical demand and generator-outage scenarios. The results highlight three key findings.

\begin{table}[t]
\centering\small
\caption{Reliability and cost by case}
\label{tab:res}
% \renewcommand{\arraystretch}{1.1}
% \resizebox{\columnwidth}{!}{%
\begin{tabular}{@{}lrrr@{}}
\toprule
Metric & No storage & Pure arbitrage & SATA \\
\midrule
EENS (MWh/day)                   & 30.37 & 24.33 & 20.89 \\
$\mathrm{CVaR}_{0.95}(U)$ (MWh) & 597.8 & 486.4 & 417.8 \\
LOLH (h/day)                     & 0.505 & 0.358 & 0.190 \\
$\mathbb{E}[C]$ (\$M/day)        & 0.823 & 0.762 & 0.721 \\
\bottomrule
\end{tabular}
% }
\vspace{-1em}
\end{table}

First, operator-dispatched SATA substantially reduces congestion-driven curtailment. Relative to no storage, SATA lowers EENS by about 31\%, cuts LOLH by about 62\%, and reduces the 95\% CVaR tail by about 30\%, while lowering expected operating cost by about 12\%. The pure-arbitrage case, by contrast, reduces EENS by about 20\% and LOLH by about 29\% using the same hardware and energy budget. The ESS delivers reliability value as a transmission asset under operator-directed dispatch, where it is held available to support the constrained corridor.

Second, the SATA benefit is most pronounced on severe contingency days. The 95\% tail of daily unserved energy is roughly twenty times the mean for each case. That tail falls by about 30\% under SATA but by only about 19\% under pure arbitrage. The operator's exposure on the worst congestion days is therefore what improves most.

\begin{figure}[ht]
\centering
\vspace{-1em}
\includegraphics[width=0.9\columnwidth]{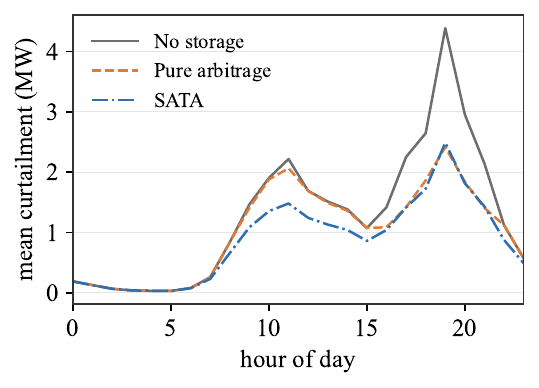}
\vspace{-1.25em}
\caption{Mean load curtailment by hour of day over all scenarios.}
\vspace{-1em}
\label{fig:curtail}
\end{figure}

Third, the cost reduction under SATA (about 12\%) is nearly double that under arbitrage (about 7\%), reflecting a shift in ESS utilization from price-driven dispatch to curtailment-avoidance and substitution for congestion-induced generation redispatch.
% Fig.~\ref{fig:ldc} shows why the congestion is structural. Without storage, the corridor operates at or near its thermal rating for a large fraction of hours, so any high-demand day or generator shortfall behind it produces curtailment. 
Fig.~\ref{fig:curtail} shows the mean curtailment by hour of day over all scenarios. The unserved energy concentrates in the high-load afternoon hours when the corridor binds. SATA removes most of this curtailment in those hours, while the committed arbitrage profile leaves a mistimed remainder.

\begin{figure}[tb]
\centering
\includegraphics[trim=0 0 0 0cm, clip, width=1\columnwidth]{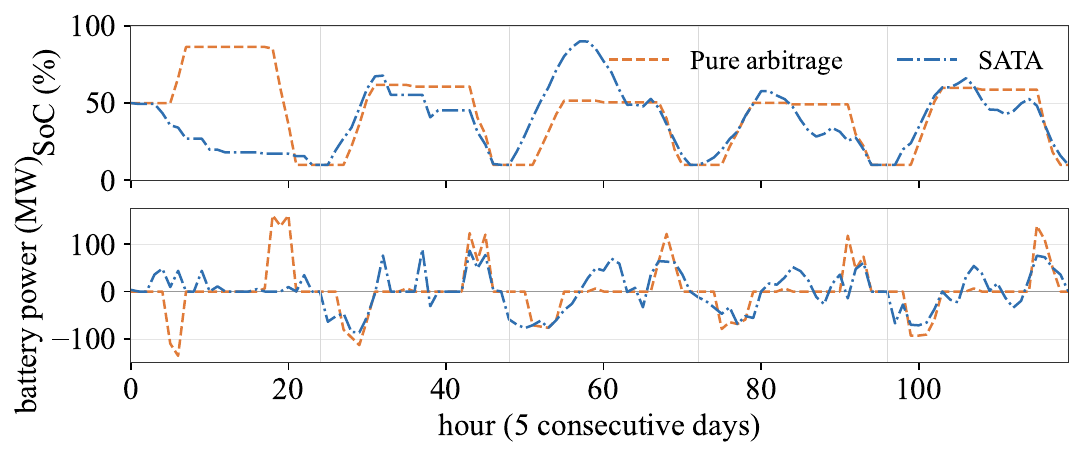}
\vspace{-1em}
\caption{State of charge (top) and ESS power (bottom) over five representative load days for the pure-arbitrage and SATA cases.}
\label{fig:soc}
\vspace{-0.5em}
\end{figure}

\begin{figure}[th]
\centering
\includegraphics[width=0.9\columnwidth]{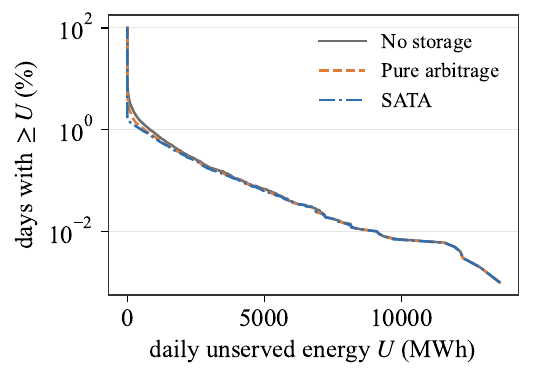}
\vspace{-1em}
\caption{Exceedance of daily unserved energy across all scenarios.}
\vspace{-1.5em}
\label{fig:exc}
\end{figure}

\begin{figure*}[t]
\centering
\includegraphics[trim=0 0 0 0cm, clip, width=0.9\textwidth]{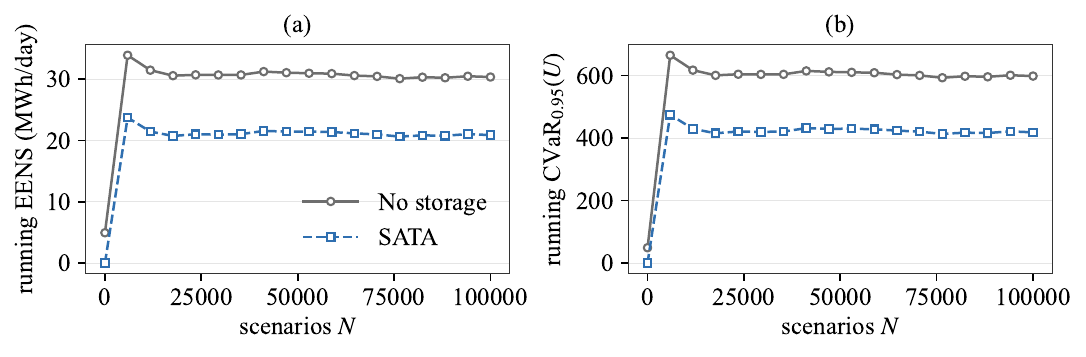}
\vspace{-1em}
\caption{Monte Carlo convergence of the running EENS and CVaR estimates across all simulated scenarios.}
\label{fig:conv}
\vspace{-1em}
\end{figure*}

Fig.~\ref{fig:soc} traces the state of charge and ESS power over five representative days ordered from light to heavy loading. The pure-arbitrage profile repeats its committed discharge cycle every day regardless of corridor conditions. The SATA schedule adapts: on a light day the ESS stays near its initial state of charge and on a heavy day it discharges fully into the binding corridor hours. Fig.~\ref{fig:exc} plots the exceedance of daily unserved energy on a logarithmic scale. The SATA curve lies below the other two across the entire tail, which confirms that the reliability improvement is largest on the worst congestion days. Fig.~\ref{fig:conv} shows the stable convergence of running EENS and CVaR.

\enlargethispage{2\baselineskip}

\subsection{Discussion}
The reliability gap between SATA and pure-arbitrage dispatch traces to a timing mismatch between the committed arbitrage schedule and the realized corridor-binding hours. Under pure arbitrage, the ESS is committed to discharge at the forecast-peak price hours. Those hours are not necessarily the hours when the corridor binds, particularly on days when a generator outage shifts the demand-generation balance behind the corridor. On such days the arbitrage resource may already be discharged when congestion peaks. The load then faces curtailment without ESS relief. SATA avoids this mismatch. The charge and discharge schedule is co-optimized against the demand and outage scenario that is realized on that day, so discharge is directed to the congested hours regardless of price signals. The CVaR result reinforces this interpretation. The worst congestion days, those on which the gap between forecast prices and realized corridor binding is most consequential, are largely unchanged by arbitrage dispatch but substantially mitigated by SATA. This observation gives concrete content to the regulatory SATA designation. The obligation to hold ESS capacity available for the transmission function, rather than commit it for energy arbitrage, is the mechanism through which the designation converts hardware into targeted corridor relief. This obligation is therefore the primary driver of the reliability difference the results reveal.
\enlargethispage{2\baselineskip}

Several modeling choices bound the scope and generality of these findings. The DC power flow model excludes reactive power and voltage constraints, and AC feasibility of the dispatch solutions is not verified. A follow-on AC security analysis would be needed before any operational implementation. Generator outages are drawn independently, so common-mode and weather-correlated failures are not represented. Correlated outages during extreme weather events would increase curtailment frequency and likely widen the reliability gap between cases. The corridor's own forced outage is excluded. Covering that risk requires a parallel reinforcement or an N-1 contingency complement to the SATA. Dual-use operation, in which the SATA earns market revenue when the corridor is not congested, is not modeled. In practice, dual-use would strengthen the economic case without necessarily reducing reliability if the operator retains dispatch priority during congested hours \cite{Twitchell2024, osti_1846604}. The congestion-price siting rule generalizes to other configurations, but the quantitative reliability findings are specific to the studied system.

%% file: Contents/conclusion.tex
This paper has demonstrated that the reliability value an energy storage system provides for corridor congestion relief is determined by its operating designation, not by its hardware. An ESS dispatched as a SATA co-optimized its schedule against the demand and outage conditions realized on each day and directed its energy to the corridor in the hours when the corridor binds. The same hardware committed to forecast arbitrage could not adapt to the realized pattern and left a material reliability and cost gap as a result. This finding gives operators and regulators a quantitative basis for the SATA designation. Co-dispatching the ESS with generation and load converts the energy budget into targeted corridor relief. The congestion-price siting rule derived here provides a reproducible, signal-driven path from interconnection screening to storage deployment.

Future work can model dual-use operation, in which the ESS earns market revenue when the corridor is not congested, to assess whether the transmission function and market participation can coexist without degrading reliability. AC verification, correlated outage models, and a parametric study across interconnection sizes and ESS durations would extend the findings to a broader class of systems and operating conditions.